\documentclass[letterpaper,twocolumn,english,floatfix, aps, pre,showpacs]{revtex4}
\usepackage[T1]{fontenc}
\usepackage[latin1]{inputenc}
\usepackage{babel}
\usepackage{graphics}

\makeatletter
\usepackage{graphicx}
\usepackage{dcolumn}
\usepackage{bm}

\makeatother
\begin{document}

\title{Precise polynomial heuristic for an NP-complete problem}

\author{P.M. Duxbury and C.W. Fay IV}

\email{duxbury@pa.msu.edu}

\affiliation{Dept. of Physics \& Astronomy, 
 Michigan State University, East Lansing, MI 48824, USA}

\begin{abstract}
We introduce a simple, efficient and precise polynomial 
heuristic for a key 
NP complete problem, minimum vertex cover.  Our method is iterative 
and operates in probability space.  Once a stable 
probability solution is found we find the 
true combinatorial solution from the probabilities.
For system sizes which are amenable to exact solution by 
conventional means, we find a correct minimum vertex cover 
for all cases which we have tested, which include random graphs 
and diluted triangular lattices of up to 100 sites.  We present precise 
data for minimum vertex cover on graphs of up to 50,000 sites.  
Extensions of the method to 
 hard core lattices gases and other NP problems are discussed.

\end{abstract}


\maketitle

There is intense  
interest in the relationships  
between statistical physics 
and computational complexity, from 
both the computer science and 
physics communities.  This activity 
has resulted in the application of 
physics methods to computer science \cite{monasson99a,weigt00a}
and clever extensions of computer 
science methods to glassy problems\cite{mezard02a}.
The NP-complete class of problems lie at the 
nexus of these dicussions. Exact solvers for 
NP-complete problems are usually restricted 
to at most a few hundred nodes which severely 
limits their practical applications.   The computational 
complexity of this class of problem has  also  
motivated a great deal of the  
interest in quantum computing, in the 
hope that this new paradigm will 
significantly improve the efficiency with which 
we can solve NP-complete problems.

In this report we introduce a new class 
of heuristic NP-complete solvers, which operate in 
probability space rather than combinatorial space.
We illustrate the potential of these methods by 
analysing the minimum vertex cover problem\cite{weigt00a,weigt01a}, which is a classical 
hard problem in the NP-complete class\cite{garey79}.  The method 
we develop is surprisingly simple and effective 
and extends in an obvious way to a broad 
class of dense packing problems  
in hard core lattice gases, which are of significant 
physical interest.
These packing problems 
are simply stated.  Given a set of 
hard core constraints, what is the 
maximum density of particles that can be 
placed on a given lattice or graph.  Minimum vertex cover 
maps to the simplest problem in this class, the  hard  
core lattice gas where only nearest 
neighbor occupation is excluded.
There is no energy parameter in the
packing problems we consider, there is only the hard core 
constraints.  Though these packing problems 
are simply stated they are proven to be in the NP class, 
and hence any significant advance in their analysis has 
broad implications in both science and technology.

The methods we introduce work by 
defining a local probability on 
each site of a graph.  In the case 
of vertex cover we intoduce 
the probability that a site has 
a guard on it.  These local probabilities 
are updated recursively using a 
relation which is locally exact for the 
probabilities.  We call this 
procedure an Exact Local 
Probability Recursion (ELoPR) algorithm.
  In the case of 
hard core lattices gases, the ELoPR  
update rule is extremely simple (see below)
and iteration of this procedure 
rapidly converges to a steady 
state occupancy probability on each site of 
a given graph.  The method is 
carried out for a given graph configuration  
and applies to any graph class, including 
random graphs, diluted regular graphs and 
graphs with structure.  This robustness 
makes ELoPR methods very attractive from 
a practical point of view.

First, we define the probability $P_i$ that 
a site, $i$, in a lattice gas is occupied by 
a particle.  If a lattice gas particle is present $P_i = 1$, 
while if the site is empty, $P_i=0$. The minimum 
vertex cover is the minimum number of ``guards'' 
which must be placed on the nodes 
of a graph so that every edge of the graph is 
covered by a guard\cite{weigt00a,weigt01a}. We define a probability 
$V_i$, so that $V_i=1$ if a guard is present,
while $V_i=0$ is a guard is absent.   We work 
with continuous probability so we also allow 
the possibility that $0<V_i<1$, which 
corresponds to degenerate sites where in some 
ground states site $i$ is occupied while in others 
it is not.  
The lattice gas and vertex cover probabilities 
are related by $V_i =  1-P_i$.   The minimum vertex 
cover corresponds to empty sites in a 
dense packing of a hardcore lattice gas with 
only nearest neighbor exclusion\cite{weigt01a}.

The ELoPR algorithm for minimum vertex cover 
is based on a simple update rule. 
 A guard is 
required at node $i$ if any of the nodes to 
which it is connected does not have a guard.
That is, the only case where a 
guard is not required is if all 
of the connected neighbors 
are already guarded.  This leads to the 
expression,
\begin{equation}
V_i = 1.0-\prod_{j=1}^{v(i)} V_{n(j)}
\end{equation}
where $i$ is the site which 
is being updated, $v(i)$ is the 
number of sites to which it is connected 
and $n(i)$ is the set of neighboring sites.
The ELoPR algorithm is consists of simply iterative
updating Eq. (1).
The computational time required for the 
minimum vertex cover is then $O(Nv_{max}n_{it})$, 
where $N$ is the number of nodes in the graph, 
$v_{max}$ is the number of neighbors of the 
most highly connected node in the graph, and 
$n_{it}$ is the number of sweeps of the lattice 
required for convergence of the site probabilities $V_i$.
We find that $n_{it}$ is at most a few thousand 
even for lattices of $50,000$ sites. 

Our implementation of the ELoPR algorithm is as follows.
We generate a graph and initialise the 
algorithm by assigning continuous random values 
of $V_i$ to each of the sites of the graph.
We then sweep through all of the sites of graph,
in a randomized order, updating $V_i$ at 
each site using Eq. (1).  We find that 
after several hundred sweeps of the lattice, 
the ELoPR procedure leads to a steady state 
value for $V_i$ on each site, for almost all finite 
initial conditions.  Remarkably, there appears
to be little metastability so that 
ELoPR usually finds a correct cover. 
However for some initial conditions, and particularly 
near the so called ``core percolation" 
threshold\cite{bauer01a} 
metastability is more likely.  However by sampling 
a set of initial conditions, usually 
only one or two are required, we are able to find the 
correct minimum vertex cover for all cases which 
we have studied. 
 
In the data presented below, we required that 
the average site probabilities, $V_i$ were converged 
to accuracy $5\times 10^{-8}$.  All of the 
calculations were carried out in double precision 
on  32-bit linux PC's.   We wrote two versions of the 
code, one in Fortran and the other in c++.  These
codes give identical results, for the same set of 
graphs, initial conditions and convergence criteria.
 We found that the 
steady state values for $V_i$ are 
either "1", "0", or an intermediate value.
This is illustrated in the top panel of Fig. 1 for a 100 node 
triangular lattice.
\begin{figure}
{\centering \resizebox*{0.95\columnwidth}{!}
{\rotatebox{-90}{\includegraphics{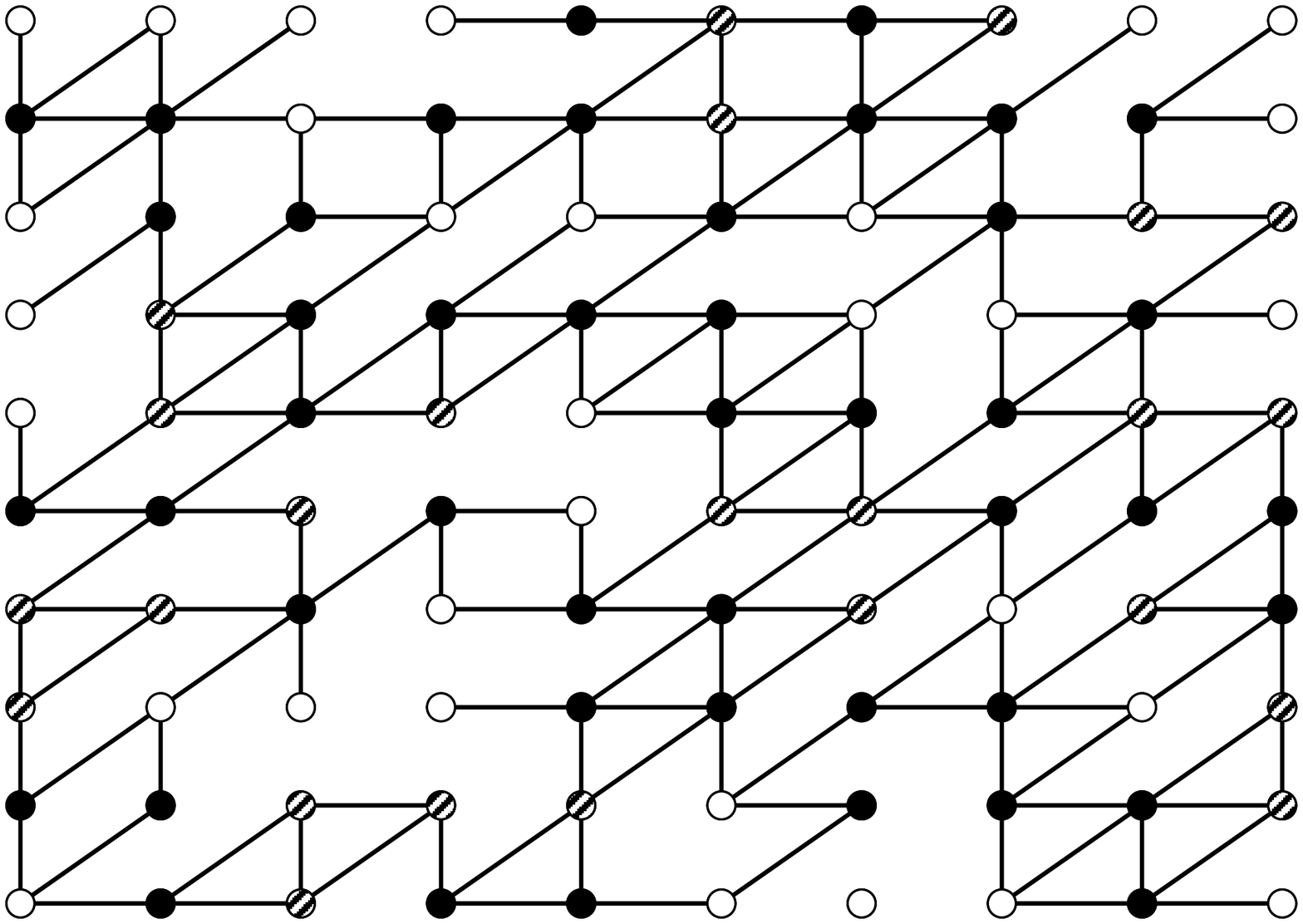}}}\par}
{\centering \resizebox*{0.95\columnwidth}{!}
{\rotatebox{-90}{\includegraphics{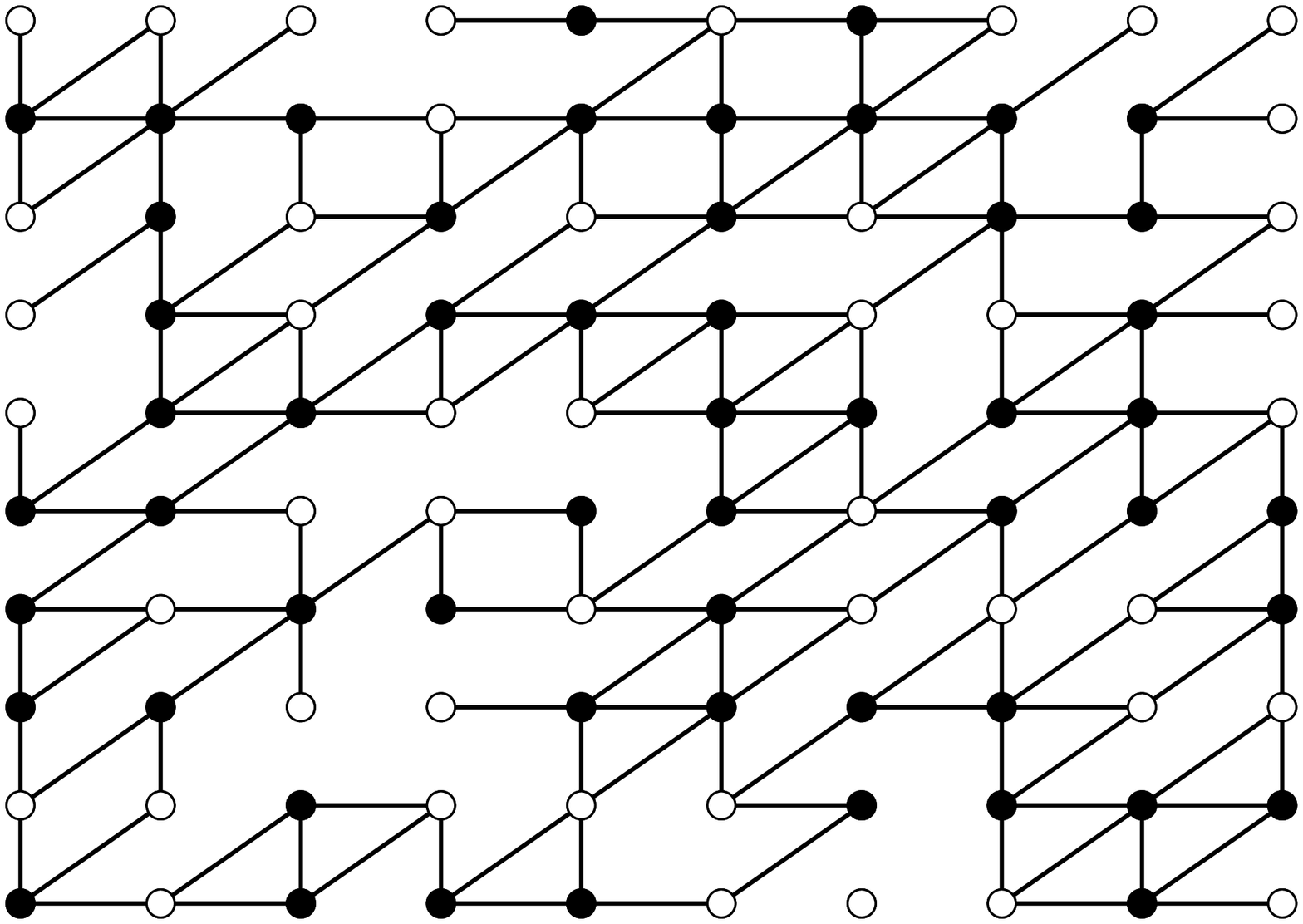}}}\par}
\caption{The minimum vertex cover on a 100 node diluted
triangular lattice. Top Figure:  The probabilistic solution
found using ELoPR.
 The solid circles are 
nodes where a guard is necessary.  The 
open circles are nodes where a guard 
is unnecessary.  The hatched nodes are degenerate. Bottom Figure: 
A specific minimum vertex cover generated from the ELoPR 
probabilities.  The minimum vertex cover for this 
graph is $54$ as was confirmed by finding the exact cover 
using an exact solver.  }
\end{figure}
The sites which have an intermediate
value are the degenerate sites, while the 
sites which have values "1" or "0" are 
the frozen sites.  We checked our algorithm 
against the exact algorithm of Aleksandar Hartmann for a large 
number of small random graphs and 
diluted triangular lattices.  In all cases, 
we found that for the lattices sizes 
accessable to exact methods 
the ELoPR procedure gives results 
which are close to exact.  The triangular 
lattice does yield some cases where ELoPR 
converges to a higher than optimal cover.
The origin of this problem is clusters of small loops 
which are common on triangular lattices, 
but not on random graphs.  The problem 
occurs in the calculation of an 
incorrect degeneracy on small loops and 
we have been able to resolve this degeneracy 
by generating a true cover from the 
ELoPR probabilities, as will be described below.

The ELoPR method for vertex cover 
is very efficient.  Finding the 
minimum vertex cover for a random graph with $N=50,000$
nodes at $c=3.0$ takes about a minute on a 
desktop linux machine.  A histogram of the 
degenerate and frozen probababilities 
for random graphs at  $c=2, 3, 4$ is presented 
in Fig. 2. 
\begin{figure}
{\centering \resizebox*{0.95\columnwidth}{!}
{\rotatebox{-90}{\includegraphics{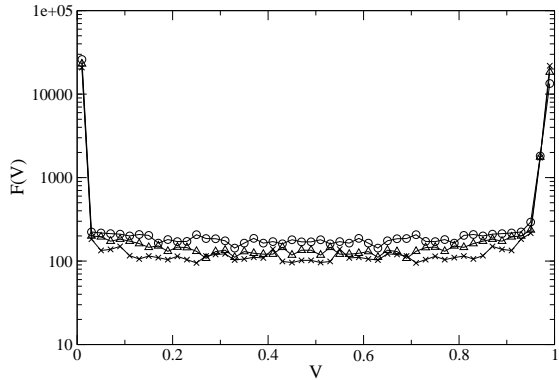}}}\par}
\caption{The distribution of vertex cover 
probabilities, $F(V)$, for $N=50,000$ site random graphs
at $c=2.0(\circ)$, $c=3.0(\Delta)$ and $c=4.0(\times)$. }
\end{figure}
The sites which are frozen covered
correspond to the delta function at one, while the 
sites which are frozen uncovered correspond to the 
delta fucntion at zero.  In addition there
is a broad, almost uniform continuum 
spread on the interval [0,1].  As the average co-ordination number of 
the graph increases the delta function at "1" increases, the 
degenerate continuum decreases and the delta function 
at "0" decreases.    
In Figure 
3, we present results for the average cover and the 
fraction of frozen sites as a function of  
bond concentration on random graphs.  These results are 
compared with data generated using survey propagation methods\cite{zhou03a},
with the replica symmetric solution and with results 
found by extrapolation using exact data on small lattices\cite{weigt00a,weigt01a}.
The replica symmetric results are believed to be a lower bound 
to the true average cover, while the survey propagation 
results\cite{zhou03a} are believed to be an improved lower bound.
It is evident from the prior results that the ELoPR results are extremely 
encouraging as they correspond to a true cover and hence 
are an upper bound to the minimum vertex cover.  If 
we accept that the survey propagation results are a lower 
bound, the true cover is tightly bounded by the combination 
of survey propagation and ELoPR. The ELoPR 
results of Fig. 3 are for one $N=50,000$ site 
random graph at each value of $c$, however at the 
resolution of this figure they are equivalent 
to the asymptotic limit ELoPR results which 
we have found by finite size scaling.
We found that the vertex cover self-averages, so 
that the results for other realisations of lattices 
of this size are identical, to the resolution of this 
figure.  The ELoPR results presented in this figure 
required about 30 minutes on a 500MHz linux machine
and includes data at 100 values of $c$ on the 
interval [0,20].    The number of frozen nodes 
found using ELoPR for a given set of 
initial conditions is higher than that 
found using exact methods, however if 
we search over a variety of initial 
conditions we find a different set of 
frozen nodes.  Moreover the frozen nodes we 
find after sampling over initial conditions are the same as the 
frozen nodes found using exact methods.

The ELoPR update formula (1) can be also be used to develop 
analytic approaches.  To illustrate this, we now reproduce
the replica symmetric result in a simple manner.
Consider the update procedure (1) on a bond-diluted Bethe lattice,
with probability $p$ that a bond is present.
We seek a steady state solution to $V $, where 
$V$ is the probability that a site far from the 
boundary of the Bethe lattice is occupied by a guard.  The 
probability that this node is occupied by a lattice gas particle is $P=1-V$.
It is most straightforward to work in terms of the 
lattice gas occupancy $P$.  We write down a recurrence 
relation for the probability that a node is occupied by 
a lattice gas particle.  If the node is part of a 
Bethe lattice of co-ordination $z$, then 
there are $\alpha = z-1$ nodes which are at a lower level in the 
tree.  We then write down a recursion relation relating $P$ 
at the current node to the values of $P$ at the $\alpha$
nodes at the lower level in the tree.   The recursion 
relation we use is Eq. (1), with $P_i=1-V_i$ and with the 
restriction that the values of $P_i$ are the same 
on all nodes, ie. we make a uniform approximation.
In order for a node to be occupied by a lattice gas 
particle, all of the nodes
to which it is connected must NOT be occupied, we 
then have, 
\begin{equation}
P = (1-pP)^{\alpha} \rightarrow e^{-cP}
\end{equation}
where the expression on the RHS is the random graph limit found
by using, $p=c/N$, $\alpha = N$, $N\rightarrow \infty$,
where $N$ is the number of nodes in the graph.   Eq. (2) is the 
branch probability.

In order to find the 
vertex cover from the branch probability $P$,  we 
take account of degeneracy which occurs when 
we connect together the $z$ branch 
probabilities at the central node of the Bethe lattice.   
If just one of the nodes to which the central  
node is connected is occupied, we 
can change its assignment so that it 
is no longer occupied while the central node  
then becomes occupied.  This can be 
done without decreasing the packing density of the 
lattice.  This is the degenerate case and 
must be included in calculating the 
average cover predicted by the Bethe lattice theory. 
The probability of 
finding this degenerate state is,
\begin{equation}
D = \alpha pP(1-pP)^{\alpha - 1} \rightarrow \alpha p P^2 \rightarrow c P^2
\end{equation}
The last expression on the RHS of Eq. (3) was 
found using Eq. (2) and then taking the random graph limit.
The minimum vertex cover is then given by,
\begin{equation}
V = 1 - P - {D\over 2} = 1 - {W(c) \over c} - {W(c)^2 \over 2c}.
\end{equation}
where $W(c) = cP$ is the Lambert function.  That 
is, the degenerate case leads to the central site 
being occupied only half of the time.  Eq. (4)  
is the replica symmetric result for the average minimum 
cover as found by
 Weigt and Hartmann\cite{weigt00a,weigt01a}.
It gives the dashed line in Fig. 3.
\begin{figure}
{\centering \resizebox*{0.95\columnwidth}{!}
{\rotatebox{-90}{\includegraphics{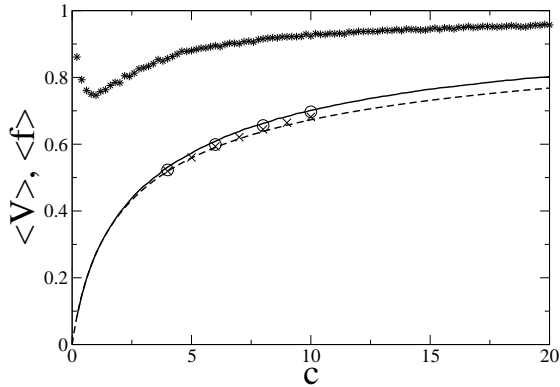}}}\par}
\caption{The minimum vertex cover <V> and the fraction of frozen sites <f>
as a function of the bond concentration, $c$, of random graphs.
 The solid line is the 
ELoPR result for  $N=50,000$ site random graphs. The 
dashed line is the replica symmetric result, from 
Eq. (4).  The circles ($\circ$) are finite size 
scaling data from Hartmann and Weigt\cite{weigt00a,weigt01a}, while the 
crosses ($\times$) are data found using the 
survey propagation algorithm\cite{zhou03a}.  The uppermost 
set of data (indicated by asterisks (*))
are the ELoPR results for the fraction of sites which 
are frozen in either the covered or uncovered state 
for one initial condition.
The remainder of the sites are degenerate and 
lead to an extensive ground state entropy.}
\end{figure}

The ELoPR method solves a combinatorial problem 
in a statistical physics sense. However in 
many cases, we also want to find
specific exact covers from these probabilities.  As seen 
in Figs. 1 and 2, the ELoPR method finds 
a relatively high fraction of the nodes to be 
either covered or uncovered.  The degenerate nodes have 
ELoPR probabilities which lie between zero and one and these values 
need to be converted into either zero or one 
in order to find a true cover.  We have developed a simple 
procedure to do this.  First we observed that the 
degenerate nodes in the ELoPR solutions 
are surrounded by covered nodes.
We identify a degenerate cluster and randomly 
choose one its nodes to uncover, ie we set $V_i=0$
on this node.  We then run ELoPR with this node fixed.
This usually removes the degeneracy of the cluster.
If it does not, we simply identify the next 
degenerate cluster and carry out the same procedure.
Carrying out this procedure to completion gives 
a true cover.  We call this procedure the 
discrete instance generator (DIG).  Once we have 
a true cover, we again calculate its minimum vertex 
cover.  A comparison of the minimum vertex cover before and 
after applying DIG to random graphs of size $N=1000$ is 
presented in Fig. 4.  It is evident that the 
DIG cover and the ELoPR cover are very close
for all values of $c$.
\begin{figure}
{\centering \resizebox*{0.95\columnwidth}{!}
{\rotatebox{-90}{\includegraphics{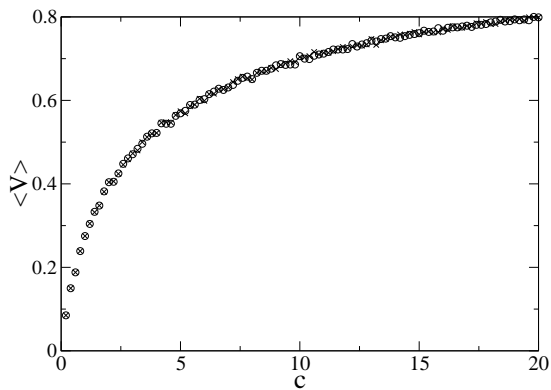}}}\par}
\caption{The minimum vertex cover before (crosses)
and after (circles) generating a true cover using DIG, for $N=1000$
site random graphs.}
\end{figure}

We also used ELoPR and DIG to find the 
 minimum vertex cover on diluted 
triangular lattices, with similarly impressive 
results. Some metastability occurs, as in the random 
graph case, however this is resolved by a 
sampling of different initial conditions, with 
the probability of finding the ground state 
with a give initial condition being well above 
50\% for all cases we have studied. 
The presence of small loops in the 
triangular lattice causes some 
deviation of the ELoPR cover from the true 
cover.  However the best ELoPR 
solution followed by the DIG procedure leads 
to an exact cover for all cases we have studied 
by conventional means, for example the $n=100$ node 
case of Fig. 1.

We are exploring many extensions and 
applications of the ELoPR method.  Firstly, 
the update procedure (1) is not restricted to 
nearest neighbors and is valid for any graph 
structure.  One interesting problem class is dense 
packing of topologically disordered graphs, such as 
voronoi tesselations of the plane.  To extend the 
method beyond hard core packing problems however, we need to be able 
to include energy parameters in the analysis, so that for example  
competing interactions may be treated.  We have 
developed an ELoPR procedure which includes  
energy terms and applies to other 
NP-complete problems, for example to 
the coloring problem. The 
update procedure is more complex, and 
includes a sum over all possible 
states of the neighboring sites.  For 
a lattice gas problem, we then 
have to sum over $2^{v(i)}$ configurations, 
even in the simplest case.  Nevertheless, this is 
still encouraging for problems having 
finite connectivity, as is the case 
for many problems of physical and technological 
interest.  Even in cases, such as coloring 
and K-SAT, where there are more degrees of 
freedom, it is possible to reduce the 
problem to $2^{v(i)}$ by using exact symmetries of the 
probabilities in the problem.  A 
presentation of these applications of the 
ELoPR concept will be presented elsewhere.\\

We acknowledge and thank Alexander Hartmann for providing 
his exact vertex cover algorithm.
This work has been supported by the DOE under contract DE-FG02-90ER45418.


\end{document}